\documentclass[showpacs,showkeys,amsmath,amssymb,preprint,11pt]{revtex4}
\usepackage{epsfig,dcolumn,bm}

\begin{document}

\title{Strangeness  $s = -3 $ dibaryons in a chiral quark model}

\author{Dai Lian-Rong$^{1}$}\thanks{E-mail: dailr@lnnu.edu.cn}
\author{Zhang Dan$^{2,3}$}\thanks{E-mail: zhangdan@mail.ihep.ac.cn}
\author{Li Chun-Ran$^{1}$}
\author{Tong Lei$^1$}
\affiliation{\small
$^1$Department of Physics, Liaoning Normal University, 116029, PR China\footnote{Mailing address.}\\
$^2$Institute of High Energy Physics, P.O. Box 918-4, Beijing 100049, PR China\\
$^3$Graduate School of the Chinese Academy of Sciences, Beijing, PR
China}

\begin{abstract}
The structures of $N\Omega_{(2,1/2)}$ and $\Delta\Omega_{(3,3/2)}$
with strangeness $s=-3$  are dynamically studied in both the chiral
SU(3) quark model and the extended chiral SU(3) quark model by
solving a resonating group method (RGM) equation.  The first model
parameters are taken from our previous work, which gave a
satisfactory description of the energies of the baryon ground
states, the binding energy of the deuteron, the nucleon-nucleon(NN)
scattering phase shifts, and the hyperon-nucleon (YN) cross
sections. The effect from the vector meson fields is very similar to
that from the one-gluon exchange interaction, both in the chiral
SU(3) quark model and the extended chiral SU(3) quark model, the
$N\Omega_{(2,1/2)}$ and $\Delta\Omega_{(3,3/2)}$ systems are wealy
bound states. The second  model parameters are also taken from our
previous work by fitting the KN scattering process.  when the mixing
of scalar mesons are considered, the $N\Omega_{(2,1/2)}$ and
$\Delta\Omega_{(3,3/2)}$ systems change into unbound state.
\end{abstract}



\maketitle

\section{Introduction}
Searching dibaryon both theoretically and experimentally has
attracted  worldwide attention since Jaffe predicted the H particle
in 1977\cite{Jaffe}. No doubt, studying dibaryon can enrich our
knowledge of the strong interaction in the short-range and can
further understand the basic theory of the strong interaction,
Quantum chromodynamics (QCD), especially its non-perturbative QCD
(NPQCD) effect. Because of complexity in NPQCD effect at the lower
energy region, one has to develop QCD-inspired models, e.g., the MIT
bag model \cite{MIT}, cloudy bag model \cite{cloudy}, Friedberg-Lee
nontopological soliton model\cite{fried}, Skyrme topological soliton
model \cite{skyrme}, the constituent quark model
\cite{consti,tubinggen}, etc. Recently, there has been a hot debate
regarding the proper effective degrees of freedom of the constituent
quark model \cite {debate}.

Among those models, the chiral $SU(3)$ quark model is one of the
most successful ones \cite{zhang97}. With such model, not only the
single baryon properties can be explained\cite{su3a}, but also the
nucleon-nucleon (NN) scattering phase shifts and the hyperon-nucleon
(YN) cross sections can be better reproduced \cite{zhang97}.  Based
on this chiral $SU(3)$ quark model, the prediction of  possible
dibaryons can be found in Refs.\cite{su3}. In Ref.\cite{dai}, we
further extended our chiral $SU(3)$ quark model to include the
coupling between the quark and vector chiral fields, named as the
extended SU(3) quark model. Such extension was made mainly based on
the following facts. Firstly, in the study of NN interactions on
quark level, the short-range feature can be explained by one gluon
exchange (OGE) interaction and the quark exchange effect, while in
the traditional one boson exchange (OBE) model on baryon level it
comes from vector meson ($\rho,\omega, K^*$ and $\phi$) exchanges.
Secondly, Glozman and Riska proposed the boson exchange model
\cite{glozman}, and found that the OGE can be replaced by
vector-meson coupling in order to elucidate baryon structure.
Therefore, in Ref. \cite{dai}, the vector meson exchange was
dynamically studied on quark level.  From our study on deuteron
structure and the $NN$ scattering phase shifts in the extended
chiral $SU(3)$ quark model, we find that quark-vector chiral field
coupling interactions can substitute the OGE mechanism on quark
level. In the extended chiral $SU(3)$ quark model, instead of the
OGE interaction, the vector meson exchanges play a dominate role in
the short range part of the quark-quark interactions. Since
geometric size of a dibaryon is small, the short range feature of
the interactions should be important in describing its structure.
Hence, a further investigation on structure of dibaryon  in the
extended chiral $SU(3)$ quark model should be helpful in resolving
this issue.

As far as the quark-quark interaction is concerned, $N\Omega$ and
$\Delta \Omega$, just like $N\phi$ state, are special systems which
are composed of two color singlet hadrons with no common flavor
quarks. In $N\Omega$ or $\Delta \Omega$ one-channel study, there is
no quark exchange between $N(\Delta)$ and $\Omega$ cluster, and no
OGE interaction. Whether $N\Omega$ or $\Delta\Omega$ can form a
bound state like $N\phi$ \cite{nfi, h73} is an interesting issue.
Moreover, in Ref. \cite{liqb}, Li and Shen gave the prediction of
binding energies of $N\Omega_{(2,1/2)}$  and
$\Delta\Omega_{(3,3/2)}$ systems in the chiral $SU(3)$ quark model,
and their results showed that $N\Omega_{(2,1/2)}$ and
$\Delta\Omega_{(3,3/2)}$ systems are the weakly bound states.
Because of the available nucleon beam and the only weak decay modes
of $\Omega$, they think the $N\Omega$ state would be easier to
search experimentally. In this work, we will further study the two
special $N\Omega_{(2,1/2)}$ and $\Delta \Omega_{(3,3/2)}$ systems in
the extended chiral $SU(3)$ quark model, in which vector meson
exchange dominates the short range interaction.  This study will
make us deeply understanding the short range quark-quark interaction
and get more knowledge of the coupling between quark and $\sigma$
chiral field.

The paper is arranged as follows. A brief introduction of the model
is outlined in section II. The results and discussions are provided
in section III. A summary is given in section IV.

\section{Formulation}

The  chiral $SU(3)$ quark model \cite{zhang97} and the extended
chiral $SU(3)$ quark model\cite{dai} has been widely described in
the literature \cite{zhang97, dai} and we refer the reader to those
works for details. Here we just give the salient feature of these
two models.

In these two models, the total Hamiltonian of baryon-baryon systems
can be written as
\begin{eqnarray}
& H & =\sum\limits_{i}T_i-T_{\rm G}+\sum\limits_{i<j}V_{ij},
\end{eqnarray}
where $\sum\limits_{i}T_i-T_{\rm G}$ is the kinetic energy of the
system, and $V_{ij}$ represent the quark-quark interactions,
\begin{eqnarray}
& V_{ij} & =V_{ij}^{\rm OGE}+ V_{ij}^{\rm conf}+V_{ij}^{\rm ch},
\end{eqnarray}
where $V_{ij}^{\rm OGE}$ is the OGE interaction, and $V_{ij}^{\rm
conf}$ is the confinement potential. $V_{ij}^{\rm ch}$ represents
the chiral fields induced effective quark-quark potential. In the
chiral $SU(3)$ quark model, $V_{ij}^{\rm ch}$ includes the scalar
boson exchanges and the pseudoscalar boson exchanges,

\begin{eqnarray} & V_{ij}^{\rm ch} & =
\sum^{8}_{a=0} V_{{\sigma}_a} ({\bf{r}_{ij}}) + \sum^{8}_{a=0}
   V_{{\pi}_a} ({\bf{r}_{ij}})
\end{eqnarray}
and in the extended chiral $SU(3)$ quark model, the vector boson
exchange  are also included,
\begin{eqnarray} & V_{ij}^{\rm ch} & =
\sum^{8}_{a=0} V_{{\sigma}_a} ({\bf{r}_{ij}}) + \sum^{8}_{a=0}
   V_{{\pi}_a} ({\bf{r}_{ij}})+ \sum^{8}_{a=0}V_{{\rho}_a}
   (\bf{r}_{ij})
\end{eqnarray}
Here $\sigma_{0},\cdots, \sigma_{8}$ are the scalar nonet fields,
$\pi_{0},\cdots, \pi_{8}$ are the pseudoscalar nonet fields and
$\rho_{0},\cdots, \rho_{8}$ are the vector nonet fields. The
expressions of these potentials can be found in the literature
\cite{dai}.

All the model parameters are taken from our previous work, which
gave a satisfactory description of the energies of the baryon ground
states, the binding energy of the deuteron, the NN scattering phase
shifts. Here we briefly give the procedure for the parameter
determination.  The three initial input parameters, i.e., the
harmonic-oscillator width parameter $b_{u}$, the up(down) quark mass
$m_{u(d)}$ and the strange quark mass  $m_{s}$, are taken to be the
usual values:  $b_{u}=0.5$ fm for the chiral SU(3) quark model and
0.45 fm for the  extended chiral SU(3) quark model $m_{u(d)}=313$
MeV, and $m_{s}=470$ MeV. The coupling constant for scalar and
pseudoscalar chiral field coupling,  $g_{ch}$, is fixed by the
relation:
\begin{eqnarray}
\frac{g_{NN\pi}^2}{4\pi} = \frac{9}{25}~ \frac{m_u^2}{M_N^2}~
\frac{g_{ch}^2}{4\pi}~,
\end{eqnarray}
with the  experimental value  ${g_{NN\pi}^2}/{4\pi}=13.67$. The
coupling constants for vector coupling of the vector-meson field is
taken to be $g_{\rm chv}=2.351$ (set I)  and $g_{\rm chv}=1.973$
(set II), respectively, the same as used in the NN case\cite{dai}.
The  masses  of the mesons  are taken to be the experimental values,
except for the  $\sigma$ meson. The $m_{\sigma}$ is adjusted to fit
the binding energy of the deuteron.  The OGE coupling constants and
the strengths of the confinement potential are fitted by baryon
masses and their stability conditions.  All the parameters are
tabulated in  Table I, where the first set is for the original
chiral SU(3) quark model, the second and third sets for the extended
chiral SU(3) quark model by taking $f_{\rm chv}/g_{\rm chv}$ as 0
and 2/3,respectively. Here $f_{\rm chv}$ is the coupling constant
for tensor coupling of the vector meson fields.

From Table I one can see that for both set II and set III, $g_{u}^2$
and $g_{s}^2$ are much smaller than the values of set I. This means
that in the  extended chiral SU(3) quark model, the coupling
constants of OGE are greatly reduced when the coupling of quarks and
vector-meson field is considered. Thus the OGE that plays an
important role of the quark-quark short-range interaction in the
original chiral SU(3) quark model is now nearly replaced by the
vector-meson exchange.  In other words, the mechanisms of the
quark-quark short-range interactions in these two models are quite
different.
\begin{table}[htb]
\begin{small}
\caption{Model parameters. Meson masses and cutoff masses:
$m_{\pi}$=138MeV, $m_{K}$=495 MeV, $m_{\eta}$=549 MeV,
$m_{\eta'}$=957 MeV, $m_{\sigma'}= m_{\epsilon}$= $m_{\kappa}$=980
MeV, $m_{\rho}$=770 MeV, $m_{K^*}$=892 MeV, $m_{\omega}$=782 MeV,
$m_{\phi}$=1020 MeV, $\Lambda$=1100 MeV for all mesons.}
\begin{center}
\begin{tabular*}{140mm}{@{\extracolsep\fill}cccc}
\hline\hline
                       & Chiral $SU(3)$&
\multicolumn{2}{c}{~~~Extended  Chiral  }   \\
                       & quark model&
\multicolumn{2}{c}{~~$SU(3)$ quark model}   \\\\
&          & ~~~~~~ set I ~~~~~&  set   II     \\
\hline
$b_u (\rm fm)$             & 0.5      & ~~~~~~ 0.45  ~~~~   & 0.45      \\
$g_{\rm NN\pi}$            & 13.67    & 13.67    & 13.67     \\
$g_{\rm ch}$               & 2.621    & 2.621    & 2.621     \\
$g_{\rm chv}$              & 0        & 2.351    & 1.973     \\
$f_{\rm chv}/g_{\rm chv}$      & 0        & 0        & 2/3       \\
$m_{\sigma}$(MeV)      & 595      & 535      & 547       \\
$g_u^2$                & 0.766    & 0.056    & 0.132     \\
$g_s^2$                & 0.846   & 0.203    & 0.250     \\
$a_{uu}$ (MeV/${\rm fm}^2$)     & 46.6     & 44.5     & 39.1      \\
$a_{us}$ (MeV/${\rm fm}^2$)     & 58.7     & 79.6     & 69.2      \\
$a_{us}$ (MeV/${\rm fm}^2$)     & 99.2     & 163.7     & 142.5      \\
$a^{0}_{uu}$ (MeV/${\rm fm}^2$) & -42.4     & -72.3     & -62.9  \\
$a^{0}_{us}$ (MeV/${\rm fm}^2$) & -36.2    & -87.6     & -74.6   \\
$a^{0}_{ss}$ (MeV/${\rm fm}^2$) & -33.8    & -108.0     & -91.0\\
\hline
                       &          &          &           \\
$B_{ \rm deu}$(MeV)         & 2.13     & 2.19     & 2.14      \\
\hline\hline
\end{tabular*}
\end{center}
\end{small}
\end{table}

In order to obtain more information of the structure of $N\Omega$
and $\Delta \Omega$ systems, parameters fitting kaon-nucleon ($KN$)
scattering process \cite{h72} are also adopted, where the scalar
meson mixing between the flavor singlet and octet mesons must be
considered.

With all parameters determined, the two baryon systems on quark
level  can be dynamically studied in the framework of the RGM, a
well established method for studying the interaction between two
clusters. The wave function of the two baryon systems is of the form
\begin{eqnarray}
\Psi_{ST}={\cal{A}}[\phi_{A}({\bf \xi_1, \xi_2}) \phi_{B}({\bf
\xi_4, \xi_5})\chi({\bf R_{AB}})],
\end{eqnarray}
where $\bf{\xi_1}$ and $\xi_2$ are the internal coordinates for the
cluster A, and $\bf \xi_4$ and $\xi_5$ are the internal coordinates
for the cluster B.~ ${\bf R_{AB}}= {\bf R_{A}}-{\bf R_{B}}$ is the
relative coordinate between the two clusters A and B. The $\phi_{A}$
and $\phi_{B}$ are the antisymmetrized internal cluster wave
function of A and B, and $\chi({\bf R_{AB}})$ the relative wave
function of the two clusters. The symbol $\cal{A}$ is the
antisymmetrizing operator defined as
\begin{eqnarray}
{\cal{A}}=1-\sum\limits_{i\in A,j\in B}P_{ij},
\end{eqnarray}
where $P_{ij}$ is the permutation operator of i-th and j-th quarks.
Expanding unknown $\chi({\bf R_{AB}})$ by employing well-defined
basis wave functions, such as Gaussian functions, one can solve the
RGM equation for a bound-state problem or a scattering one to obtain
the binding energy or scattering phase shifts for the two-cluster
system. The details of solving the RGM equation can be found in
Refs.\cite{tang,kamimura,oka}.

\section{results and Discussion}
As mentioned above, the $N\Omega$ and  $\Delta\Omega$ systems are
very speical two-baryon states, since there is no OGE interaction
between these two clusters. In these two sysytems, the expectation
values of the antisymmetrizer in the spin-flavor-color space is
equal to 1. It means that there is no quark effect, and whether the
system is bound merely depends on the characteristics of the
interaction induced by the chiral-quark fields coupling. In this
work, we will give the results by fitting NN and KN scattering
process both in the chiral SU(3) quark model and in the extended
chiral SU(3) quark model , respectively.

\vskip .3cm \noindent \textbf{ 1. Fit the NN scattering process}
\vskip.2cm

\begin{figure}
\epsfig{file=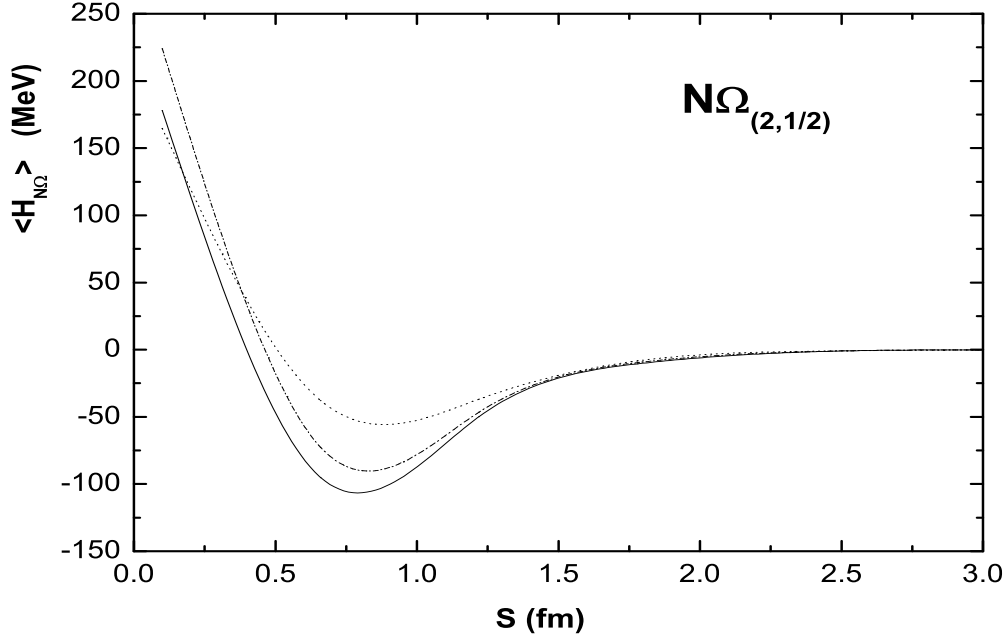,width=15.0cm,height=10.0cm} \vglue -0.5cm
\caption{\small \label{Fig.1} The GCM matrix elements of the
Hamiltonian for $N\Omega_{(2,1/2)}$ system. The dotted line
represents the results obtained in original chiral $SU(3)$ quark
model, and the solid and dash-dotted lines represent the results in
extended chiral $SU(3)$ quark model with set I and set II,
respectively.}
\end{figure}

\begin{figure}
\epsfig{file=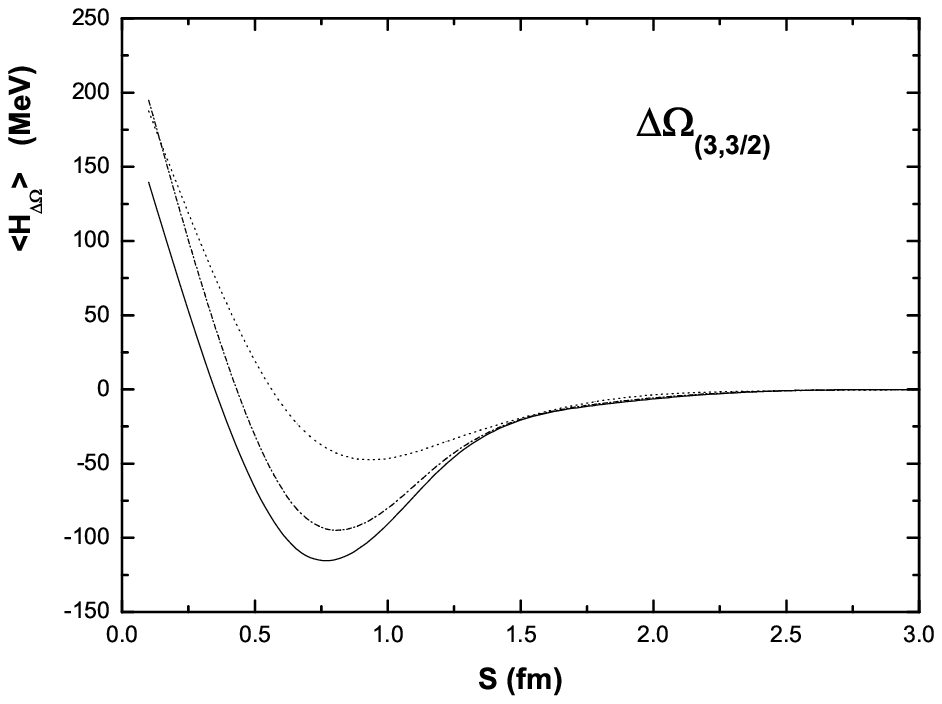,width=15.0cm,height=10.0cm} \vglue -0.5cm
\caption{\small \label{Fig.2} TThe GCM matrix elements of the
Hamiltonian for $\Delta\Omega_{(3,3/2)}$ system. The dotted line
represents the results obtained in original chiral $SU(3)$ quark
model, and the solid and dash-dotted lines represent the results in
extended chiral $SU(3)$ quark model with set I and set II,
respectively.}
\end{figure}

\begin{table}
\caption{Binding energy $B$ and rms $\overline{R}$ of the
$N\Omega_{(2,1/2)}$ and $\Delta\Omega_{(3,3/2)}$ dibaryons.
~~$B=-(E_{AB}-M_{A}-M_{B})$,~ $\overline{R}=\sqrt{\langle\,r^2
\rangle}$.}
\begin{small}
\begin{center}
\begin{tabular*}{140mm}{@{\extracolsep\fill}lllccc}
\hline\hline &&&$N\Omega_{(2,1/2)}$&&$\Delta\Omega_{(3,3/2)}$ \\
\hline
Chiral SU(3) &&$B$(MeV)&3.0&&0.6\\
 quark model&&$\overline{R}$(fm)&1.2&&1.3\\
 \hline
Extended &~ set I~ &$B$(MeV)&20.4&&25.6\\
 chiral SU(3) &&$\overline{R}$(fm)&0.9&&0.9\\
quark model&~ set II~ &$B$(MeV)&12.1&&14.4\\
&&$\overline{R}$(fm)&1.0&&0.9\\
\hline\hline
\end{tabular*}
\end{center}
\end{small}
\end{table}

\begin{table}
\caption{Contributions of various terms to binding energy for
$N\Omega_{(2,1/2)}$ dibaryon, the unit for energy is in MeV.}
\begin{small}
\begin{center}
\begin{tabular*}{140mm}{@{\extracolsep\fill}cccc}
\hline\hline
                       & Chiral $SU(3)$&
\multicolumn{2}{c}{~~~Extended  Chiral  }   \\
                       & quark model&
\multicolumn{2}{c}{~~$SU(3)$ quark model}   \\
&          & ~~~~~~ set I ~~~~~&  set   II     \\
\hline
$Kine$             & 23.1      & 44.7  & 34.4      \\
$OGE$            & 0    & 0   & 0     \\
$\pi$               & 0    & 0   &0    \\
$K$              & 6.0       & 13.4   & 9.8     \\
$\eta$      & 0.9       & 2.2        & 1.5      \\
$\eta'$      & -0.6     & -1.6      & -1.1       \\
$\sigma$     & -55.0    & -121.4   & -93.2    \\
$\sigma'$     &0   & 0    & 0     \\
$\kappa$     & 6.4    & 12.9    & 9.5     \\
$\epsilon$     & 15.8     &31.3     & 24.2     \\
$\rho$     &     & 0    & 0    \\
$K^*$     &     & -1.1    & 1.9 \\
$\omega$     &   & 0     & 0   \\
$\phi$  &     & 0    & 0\\ \hline\hline
\end{tabular*}
\end{center}
\end{small}
\end{table}

\begin{table}
\caption{Contributions of various terms to binding energy for
$\Delta\Omega_{(3,3/2)}$ dibaryon, the unit for energy is in MeV.}
\begin{small}
\begin{center}
\begin{tabular*}{140mm}{@{\extracolsep\fill}cccc}
\hline\hline
                       & Chiral $SU(3)$&
\multicolumn{2}{c}{~~~Extended  Chiral  }   \\
                       & quark model&
\multicolumn{2}{c}{~~$SU(3)$ quark model}   \\
&          & ~~~~~~ set I ~~~~~&  set   II     \\
\hline
$Kine$      & 19.7     & 53.7  & 39.3      \\
$OGE$       & 0    & 0   & 0     \\
$\pi$       & 0    & 0   &0    \\
$K$         & 3.6       & 12.6   & 8.8     \\
$\eta$      & 1.9       & 8.6       & 5.6     \\
$\eta'$     & -1.3     & -6.6      & -4.1       \\
$\sigma$    & -46.2  & -141.8   & -105.9    \\
$\sigma'$   & 0   & 0    & 0     \\
$\kappa$    & 7.8   & 24.0   & 16.9    \\
$\epsilon$  & 13.1     &37.4     & 27.9    \\
$\rho$      &     &   0  & 0    \\
$K^*$       &     & -11.5    & -3.0 \\
$\omega$    &   & 0     & 0   \\
$\phi$      &   & 0    & 0\\ \hline\hline
\end{tabular*}
\end{center}
\end{small}
\end{table}

Firstly, the same as  $N \phi$ state \cite{h73}, the model
parameters, listed in Table I, are taken from our previous work,
which gave a satisfactory description of the energies of the baryon
ground states, the binding energy of the deuteron, the NN scattering
phase shifts. The parameters in the extended chiral $SU(3)$ quark
model for two different cases are shown, one is no tensor coupling
of the vector mesons with $f_{\rm chv}/g_{\rm chv}=0$ (set I),
another involves tensor coupling of the vector mesons with $f_{\rm
chv}/g_{\rm chv}=2/3$ (set II). From the Table I, we can see that,
compared to chiral $SU(3)$ quark model, the coupling constants
$g_u^{2}$ and $g_s^{2}$ of the OGE for both set I and set II cases,
were greatly reduced when the vector meson field coupling is
considered, which manifests that the OGE interaction is quite weak
in the extended chiral $SU(3)$ quark model. Instead of the OGE, the
vector meson exchanges play dominate role in the short range part of
the interaction between two quarks, so that the mechanism of the
quark-quark short range interaction of the two models is totally
different. The quark-quark short range interaction is from the OGE
in the chiral $SU(3)$ quark model, while it is mainly from vector
meson exchanges in the extended chiral $SU(3)$ quark model.

As mentioned above, the $N\Omega$  and $\Delta\Omega$ are very
special two-baryon states since these two color singlet clusters
have no common flavor quarks.  Here we study the $N\Omega$ and
$\Delta \Omega$ states by treating $N(\Delta)$ and $\Omega$ as two
clusters and solving the corresponding RGM equation.  Fig.1 shows
the diagonal matrix elements of the Hamiltonian for the $N\Omega$
system ($<H_{N\Omega}>$) in the generator coordinate method (GCM)
\cite{tang} calculation, which includes the kinetic energy of the
relative motion and the effective potential between  $N$ and
$\Omega$ and can be regarded as the effective Hamiltonian of two
clusters  $N$ and $\Omega$ qualitatively. $s$  denotes the generator
coordinate which can qualitatively describe the distance between the
two clusters.  Similarly, Fig.2 shows the GCM diagonal matrix
elements of the Hamiltonian for the $\Delta\Omega$ system
($<H_{\Delta\Omega}>$). From Fig.1 and Fig.2, one can see that,
there exists the medium range attractive interaction for $N\Omega$
and $\Delta\Omega$ systems. From our analysis, the attraction
dominantly comes from the $\sigma$ field coupling which is spin and
flavor independent. To study whether such an attraction can make for
a bound state of the $N\Omega$ and $\Delta\Omega$ systems, we solve
the RGM equation for the bound state problem. The calculated binding
energies and corresponding root-mean-square radii (RMS) of $N\Omega$
and $\Delta\Omega$ systems are tabulated in Table II.  It is shown
that in the original chiral $SU(3)$ quark model, we get weakly bound
states of $N\Omega$ with 3.0 MeV  binding energy and $\Delta\Omega$
with 0.6 MeV, and the  corresponding RMS are 1.2 fm and 1.3 fm,
respectively.

The results of the extended chiral $SU(3)$ quark model for two
different cases are shown. In set I, i.e., there is  no tensor
coupling, the prediction of binding energy is 20.4 Mev for $N\Omega$
state and 25.6 Mev for $\Delta\Omega$ state. In set II, i.e., tensor
coupling is involved, the prediction of binding energy is 12.1 Mev
for $N\Omega$ state and 14.4 Mev for $\Delta\Omega$ state. Actually,
as can be seen in Fig.1 and Fig.2, the $N\Omega$ and $\Delta\Omega$
interactions in set I are more attractive than  those other two
cases, thus we get a relatively larger binding energies for
$N\Omega$ and $\Delta\Omega$ states. Anyway, no matter whether the
OGE or the vector meson exchange controls the quark-quark short
range interaction, the main properties of $N\Omega$ and
$\Delta\Omega$ keep unaffected, i.e., the  $N\Omega$ and
$\Delta\Omega$ systems are all not deeply bound states.

In Table III and Table IV, contributions of various terms to binding
energies for $N\Omega$ and $\Delta\Omega$ dibaryons are given.  We
can see, in the chiral $SU(3)$ quark model, there is no contribution
from OGE, while the $\sigma$ exchange dominantly provides the
attractive interaction for $N\Omega_{(2,1/2)}$ and
$\Delta\Omega_{(3,3/2)}$ systems.  In the extended chiral $SU(3)$
quark model, there is no contribution from $\rho,\omega$ and $\phi$
exchanges and little contribution from $K^{\ast}$ exchange, and the
attraction in these two special system also dominantly  comes from
$\sigma$ exchange. Simultaneously, more attraction from $\sigma$
meson can be obtained in extended chiral $SU(3)$ quark model, the
same as $N \phi$ state \cite{h73}. In our calculation, the model
parameters are fitted by the NN scattering phase shifts, and the
mass of $\sigma$ is adjusted by fitting the  binding energy of
deuteron, thus the value of $m_\sigma$ is somewhat different for
these three cases. In set I of the extended chiral $SU(3)$ quark
model, the mass of the $\sigma$ meson is smaller than other two
cases, thus $N\Omega$ and $\Delta\Omega$ can get more attraction and
more binding energies are obtained.

\vskip .3cm \noindent \textbf{ 2. Fit the KN process} \vskip .2cm

Recently,  both the chiral SU(3) quark model and the extended chiral
SU(3) quark model have been extended from the study of baryon-baryon
scattering processes to the baryon-meson systems \cite{hf} by
solving a resonating group method (RGM) equation. In order to study
the kaon-nucleon (KN) scattering, the scalar meson mixing between
the flavor singlet and octet mesons must be considered to explain
the experimental phase shift. Thus, it is interesting to see what
would  $N\Omega$ and $\Delta\Omega$ systems become  after
considering the scalar meson mixing for these two special systems?
In our calculation, scalar $\sigma$, $\epsilon$ mesons are mixed
from $\sigma_{0}$ and $\sigma_{8}$ with
\begin{eqnarray}
\sigma =\sigma_{8}\sin\theta^{S}+\sigma_{0}\cos\theta^{S},\nonumber\\
\epsilon~ =\sigma_{8}\cos\theta^{S}-\sigma_{0}\sin\theta^{S},
\end{eqnarray}
The mixing angle $\theta^{S}$ has been an open issue because the
structure of the $\sigma$ meson is still unclear and controversial.
Here we adopt two possible values as did in \cite {hf,h72}. One is
the ideal mixing with $\theta^{S}= 35.264^\circ$. This is an extreme
case in which the $\sigma$ exchange may occur only between u(d)
quarks, while $\epsilon$ occurs between s quarks. Another mixing
angle with $\theta^{S}= -18^\circ$ adopted was provided by Dai and
Wu based on their investigation of a dynamically spontaneous
symmetry breaking mechanism \cite{daiyb}.  when we adopt the model
parameters which are taken from Ref.\cite{h72} by fitting KN
scattering processes, the calculated results show  that both
$N\Omega_{(2,1/2)}$ and $\Delta\Omega_{(3,3/2)}$ systems become
unbound states for both ideally mixing and $\theta^{S}= -18^\circ$.
We make an analysis for the $N\Omega$ state. In first case, the
mixing of scalar meson is taken to be the ideally mixing, the
attraction from the $\sigma$ meson has reduced to zero. In second
case, $\theta^{S}= -18^\circ$, the attraction of the $\sigma$ meson
can be reduced a lot, thus the $N\Omega$ become unbound state.  For
$\Delta\Omega$ state, it is the same as the $N\Omega$ state, the
attraction from the $\sigma$ meson has reduced to zero with the
ideally mixing and the attraction of the $\sigma$ meson can be
reduced a lot with $\theta^{S}= -18^\circ$. The results tell us that
the mixing $\theta^{S}$ has large effect on the structure of
$N\Omega$ and $\Delta\Omega$ dibaryons.   For weakly bound system,
the structure will be changed from bound state to unbound state when
the mixing $\theta^{S}$ is considered.

Here we would like to point out that when the mixing angle of scalar
meson is considered, the parameters are obtained by fitting KN
scattering processes, not by fitting NN and YN scattering processes.
At the moment, we could not get a set of unified parameter to fit
all of the scattering data of KN, NN and YN systems. However, we
would like to see the effect on the structure of $N\Omega$ and
$\Delta\Omega$  from various set of parameters. The results tell us
that different sets of parameters have  some effect  on the
structure of $N\Omega$ and $\Delta\Omega$  dibaryons, and the
binding energies would changed with different sets of parameters.
For these two weakly bound systems, the structure property could be
changed from bound state to unbound state.

\section{summary}
In this work, we dynamically study the structure of $N\Omega$ and
$\Delta\Omega$ dibaryons with strangeness $s=-3$ in the chiral
$SU(3)$ quark model as well as in the extended chiral $SU(3)$ quark
model by solving the RGM equation. All the model parameters are
taken from our previous work. Firstly, we adopt the  model
parameters which can give a satisfactory description of the energies
of the baryon ground states, the binding energy of the deuteron, and
the NN scattering phase shifts. The calculated results show that,
for $N\Omega$ and $\Delta\Omega$ systems, the effect from the vector
meson fields is somewhat similar to that from OGE interaction.  The
$N\Omega$ and $\Delta\Omega$ dibaryons are still weakly  bound state
when the vector meson exchanges control the short range part of the
quark-quark interaction. Secondly, we take the  model parameters
with the mixing of scalar mesons by fitting KN scattering phase
shifts. The result shows that the $N\Omega$ and $\Delta\Omega$
systems would become unbound both in the chiral $SU(3)$ quark model
and in the extended chiral $SU(3)$ quark model. It should be noted
that our current analysis of $N\Omega$ and $\Delta\Omega$ systems
are based on the results from the fit to KN scattering processes, in
which the scalar meson mixing must be considered.

There are some works on s=-3 dibaryons.  The  $N\Omega$ and
$\Delta\Omega$ states  was also predicted by \cite{wf}, and they
think these two states are deeply bound states. Here in this work,
we've got the different results with the Ref.\cite{wf}.
Experimentally, whether there exist the $N\Omega$ and $\Delta\Omega$
bound states can help us to deeply understand  the quark-quark
interaction.

\vspace{1.0cm} \noindent {\bf Acknowledgement} We are in debt to
Prof. Zhang Zong-ye and Prof. Yu You-wen for their suggestive and
helpful discussions. Project supported from the National Natural
Science Foundation of China (10475087; 10575047).

{}


\footnotesize
\begin{thebibliography}{}
\bibitem{Jaffe} R.~L.~Jaffe,  Phys. Rev. Lett. {\bf 38}, 195 (1977).
\bibitem{MIT} A. Chodos et al., Phys. Rev. {\bf D 9}, 3471 (1974).
\bibitem{cloudy} A. W. thomas, Adv. Nucl. Phys. {\bf 13},1(1984);\\
                 T. DeGrand et al, Phys. Rev. {\bf D 12}, 2060 (1975).
\bibitem{fried} R. Friedberg and T. D. Lee, Phys. Rev. {\bf D 15}, 1694 (1977); {\bf D 16}, 1096 (1977);{\bf D 18}, 2623 (1978).
\bibitem{skyrme} T. H. R. Skyrme, Nucl. Phys. {\bf 31},556(1962);\\
                 E. Witten,  Nucl. Phys. {\bf B160},57(1979);\\
                 G. S. Adkins, C.R. Nappi and E. Witten, ibid.  {\bf B228}, 552(1983).
\bibitem{consti} A. De Rujula, H. Georgi, and S. L. Glashow, Phys. Rev. {\bf D12}, 147(1975);
                 N. Isgur and G. Karl, ibid. {\bf 18}, 4187(1978); {\bf 19}, 2653(1979); {\bf 20}, 1191(1979).
\bibitem{tubinggen} M. Oka, K. Yazaki, Prog. Theor. Phys. {\bf 66}, 556 (1981);\\
                   A. Faessler, et al, Phys. Lett. {\bf B112}, 201 (1982);\\
                   A. Faessler, F. Fernandez, G. L\"{u}beck, K. Shimizu, Nucl. Phys. {\bf A402}, 555 (1983);\\
                   Z. Y. Zhang, K. Br\"{a}uer, A. Faessler, K. Shimizu, Nucl. Phys. {\bf A443}, 557 (1985);\\
                   F. Fernandez, A.Valcarce, U. Straub, A. Faessler, J. Phys. {\bf G10}, 2013 (1993).
\bibitem{debate} N. Isgur, Phys. Rev. {\bf D61}, 118501(2000);\\
                 N. Isgur, Phys. Rev. {\bf D62}, 054026 (2000);\\
                 L. Ya. Glozman, nucl-th/9909021;  H. Collins and H. Georgi, Phys. Rev. {\bf D59}, 094010(1999);\\
                 K. F. Liu, S. J. Dong, T. Draper,  J. Sloan, W. Wilcox, and R. M. Woloshyn, Phys. Rev. {\bf D61}, 118502(2000).
\bibitem{zhang97}  Z.~Y.~Zhang, Y.~W.~Yu, P.~N.~Shen, L.~R.~Dai, A.~Faessler and U.~Straub,  Nucl. Phys. {\bf A625} 59 (1997);
                  S. Yang, P. N. Shen, Z.~Y.~Zhang, Y.~W.~Yu, Nucl. Phys. {\bf A635} 146 (1998).
\bibitem{su3a} P.N.Shen, Y. B. Dong, Y. W. Yu, Z. Y. Zhang and T. S. H. Lee, Phys. Rev. {\bf C55}, 204(1997);\\
               H. Chen and Z. Y. Zhang, High Ener. Phys. Nucl. Phys. {\bf 20}, 937(1996).
\bibitem{su3} P. N. Shen, Z. Y. Zhang, Y. W. Yu, X. Q. Yuan,  S. Yang, J. Phys. {\bf G25},1807(1999);\\
              X. Q. Yuan, Z.~Y.~Zhang, Y.~W.~Yu, P.~N.~Shen, Phys. Rev. \textbf{C60}, 045203(1999);\\
              Z. Y. Zhang, Y.W. Yu, C. R. Ching, T. H. Ho, Z. D. Lu, Phys. Rev. \textbf{C61}, 065204(2000);\\
              Q. B. Li, P. N. Shen, Z. Y. Zhang, Y. W. Yu, Nucl. Phys.{\bf A683}, 487(2001);\\
              Q. B. Li and P. N. Shen, Phys. Rev. \textbf{C62}, 028202(2000); J. Phys. {\bf G26},1207(2000).
\bibitem{dai} L. R. Dai, Z. Y. Zhang, Y. W. Yu, P. Wang, Nucl. Phys. {\bf A727}, 321(2003).
\bibitem{glozman} L. Ya. Glozman and D. O. Riska, Phys. Reports, {\bf 268}, 1996, 263;\\
                  L. Ya. Glozman, Nucl. Phys. {\bf A663}, 103c(2000).
\bibitem{nfi} H. Gao, T.-S. H. Lee and V. , Marinov, Phys. Rev. \textbf{C63}, 022201 (R) (2001).
\bibitem{h73} F. Huang and Z. Y. Zhang, Y. W. Yu, Phys. Rev. \textbf{C73}, 025207(2006).
\bibitem{liqb} Q. B. Li and P.~N.~Shen, Eur. Phys. J. {\bf A8}, 417(2000).
\bibitem{h72} F. Huang, Z. Y. Zhang, Phys. Rev. \textbf{C72}, 024003 (2005).
\bibitem{tang}K. Wildermuth and Y. C. Tang, {A Unified Theory of the Nucleus} (Vieweg, Braunschweig, 1977).
\bibitem{kamimura}M. Kamimura, SUppl. Prog. Theor. Phys.62,236(1977).
\bibitem{oka} M. Oka and K. Yazaki, Prog. Theor. Phys.66,556(1981).
\bibitem{daiyb} Y. B. Dai and Y. L. Wu, Eur. Phys. J. C {\bf 39}, S1(2005).
\bibitem{hf} F. Huang, Z. Y. Zhang and Y. W. Yu, Phys. Rev. \textbf{C70}, 044004 (2004); \\
             F. Huang and Z. Y. Zhang, Phys. Rev. \textbf{C70}, 064004(2004); \textbf{C72}, 068201(2005);\\
             D. Zhang, F. Huang, Z. Y. Zhang, Y. W. Yu, Nucl. Phys. {\bf A756}, 215 (2005);\\
             F. Huang, D. Zhang, Z. Y. Zhang, Y. W. Yu, Phys. Rev. \textbf{C71}, 064001(2005).
\bibitem{wf} H. R. Pang, J. L. Ping, F. Wang, T. Goldman and E. G. Zhao, Phys. Rev. \textbf{C69}, 065207 (2004);\\
             F. Wang, J. L. Ping, G. H. Wu, L. J. Teng and T. Goldman, Phys. Rev. \textbf{C51}, 3411 (1995);\\
             T. Goldman, K. Maltman, G. J. Stephenson, K. E. Schmidt and Fan Wang, Phys. Rev. Lett. 59, 627(1987).
\end{thebibliography}
\end{document}